\documentclass[floatfix,aps,twocolumn,showpacs,superscriptaddress,amsmath,amstex,
amssymb,longbibliography]{revtex4-2}

\usepackage[autostyle=true]{csquotes}

\usepackage{times}
\usepackage{xcolor}
\usepackage{mathtools}
\usepackage{textcomp}
\usepackage{gensymb}
\usepackage{graphicx}
\usepackage{siunitx}
\usepackage{ulem}
\usepackage{physics}
\usepackage{appendix}
\usepackage{verbatim}

\usepackage[unicode=true,
 bookmarks=true,bookmarksnumbered=true,bookmarksopen=true,bookmarksopenlevel=2,
 breaklinks=false,pdfborder={0 0 1},backref=false,colorlinks=true]
 {hyperref}

 \hypersetup{linkcolor=blue, citecolor=blue, urlcolor=blue, filecolor=blue, pdfpagelayout=OneColumn,
  pdfnewwindow=true, pdfstartview=XYZ, plainpages=false}

\usepackage[all]{hypcap}
\usepackage{color}

\makeatletter
\def\mathcolor#1#{\@mathcolor{#1}}%
\def\@mathcolor#1#2#3{%
  \protect\leavevmode%
  \begingroup\color#1{#2}#3\endgroup%
}%
\newcommand{\msout}[1]{\text{\sout{\ensuremath{#1}}}}%

\newcommand{\sam}[2]{%
\ifmmode%
  \msout{#1}\mathcolor{red}{#2}%
\else%
  \sout{#1}\textcolor{red}{#2}%
\fi}
\newcommand{\samO}[2]{%
\ifmmode%
  \msout{#1}\mathcolor{magenta}{#2}%
\else%
  \sout{#1}\textcolor{magenta}{#2}%
\fi}

\newcommand{\cyr}[2]{%
\ifmmode%
  \msout{#1}\mathcolor{red}{#2}%
\else%
  \sout{#1}\textcolor{red}{#2}%
\fi}

\newcommand{\gab}[2]{%
\ifmmode%
  \msout{#1}\mathcolor{green}{#2}%
\else%
  \sout{#1}\textcolor{green}{#2}%
\fi}

\makeatother

\makeatletter

\newcommand\Autoref[1]{\@first@ref#1,@}
\def\@throw@dot#1.#2@{#1}
\def\@set@refname#1{
    \edef\@tmp{\getrefbykeydefault{#1}{anchor}{}}%
    \xdef\@tmp{\expandafter\@throw@dot\@tmp.@}%
    \ltx@IfUndefined{\@tmp autorefnameplural}%
         {\def\@refname{\@nameuse{\@tmp autorefname}s}}%
         {\def\@refname{\@nameuse{\@tmp autorefnameplural}}}%
}
\def\@first@ref#1,#2{%
  \ifx#2@\autoref{#1}\let\@nextref\@gobble
  \else%
    \@set@refname{#1}
    \@refname~\ref{#1}
    \let\@nextref\@next@ref
  \fi%
  \@nextref#2%
}
\def\@next@ref#1,#2{%
   \ifx#2@ and~\ref{#1}\let\@nextref\@gobble
   \else, \ref{#1}
   \fi%
   \@nextref#2%
}

\makeatother

\begin{document}

\def\be{\begin{equation}}
\def\ee{\end{equation}}

\def\myvec#1{{\bf #1}}
\def\Esw{\myvec{E}_{sw}}
\def\Hsw{\myvec{H}_{sw}}

\title{Quantum-based solution of time-dependent complex Riccati equations}

\author{D. Mart\'inez-Tibaduiza}
\email{Correspondence to: danielmartinezt@gmail.com}
\affiliation{Instituto de F\'isica, Universidade Federal Fluminense, \\ 
 Avenida Litor\^{a}nea, 24210-346 Niteroi, RJ, Brazil}
\author{C.~Gonz\'alez-Arciniegas}
\affiliation{ Department of Physics, University of Virginia, \\ 
 400714 Virginia, U.S.A}
\author{C.~Farina}
\affiliation{Instituto de F\'isica, Universidade Federal do Rio de Janeiro, \\ 
21941-972 Rio de Janeiro, RJ, Brazil}
\author{A. Cavalcanti-Duriez}
\affiliation{Instituto de F\'isica, Universidade Federal Fluminense, \\ 
 Avenida Litor\^{a}nea, 24210-346 Niteroi, RJ, Brazil}
\author{A.~Z.~Khoury}
\affiliation{Instituto de F\'isica, Universidade Federal Fluminense, \\ 
 Avenida Litor\^{a}nea, 24210-346 Niteroi, RJ, Brazil}


\begin{abstract}

Using the Wei-Norman theory we obtain a time-dependent complex Riccati equation (TDCRE) as the solution of the time evolution operator (TEO) of quantum systems described by time-dependent (TD) Hamiltonians that are linear combinations of the generators of the $\mathfrak{su}(1,1)$, $\mathfrak{su}(2)$ and $\mathfrak{so}(2,1)$ Lie algebras. 
Using a recently developed  
solution for the time evolution of these quantum systems we solve the TDCRE recursively as generalized continued fractions, which are optimal for numerical implementations, and  
establish the necessary and sufficient conditions for the unitarity of the TEO in the factorized representation.
The inherited symmetries of quantum systems can be recognized by a simple inspection of the TDCRE, allowing effective quantum Hamiltonians to be associated with it,  
as we show for the Bloch-Riccati equation whose  Hamiltonian corresponds to that of a generic TD system of the Lie algebra $\mathfrak{su}(2)$. As an application, but also as a consistency test, we compare our solution with the analytic  one for the Bloch-Riccati equation considering the Rabi frequency driven by a complex hyperbolic secant pulse generating spin inversion, showing an excellent agreement.

\end{abstract}

\maketitle

Symmetries always had an important place in physics, and they became mainstays since Emmy Noether's theorem \cite{Noether-1918}, 
where they were formally connected with conserved quantities. 
This theorem arises from the study of a lagrangian under the action of groups of infinitesimals transformations  
known as Lie groups \cite{Kosmann-Schwarzbach-2010}, which are of special interest in physics because 
they are continuous groups with the structure of a differential manifold \cite{GILMORE-2012}. 
Lie groups can be introduced through their corresponding Lie algebras \cite{van_der_Waerden_1985} with the group structures identified
from the commutation relations satisfied by the generators of the algebra. A paradigmatic example of algebraic methods, 
\textit{i.e.}, methods that use the algebraic structure to describe and solve 
physical systems, can be found in one of the many ways of solving the quantum harmonic oscillator, where ladder operators 
are introduced to diagonalize the Hamiltonian allowing a precise and elegant way of finding the 
corresponding energy levels and energy eigenfunctions \cite{Sakurai-2011}. 
Algebraic methods are important not just in the obtainment of the energy spectrum of physical systems \cite{Andrianov_1984}, 
but also in the computation of dynamical properties as the time evolution operator (TEO), Feynman propagators or Green functions 
\cite{Wilcox-1967, Dong_2007}. 
Moreover, these methods can be used in the treatment of physical systems described by time-dependent (TD) Hamiltonians, 
which are natural scenarios for describing interactions with external agents.
As a remarkable example, the so-called \textit{Wei-Norman theory} \cite{Wei_1963, Wei_1964} 
allows to find the exact solution of these systems when their Hamiltonians can be written as a 
linear combination of time-independent generators of a finite Lie algebra. 
Using this method, the Schr\"odinger equation is mapped on a set of coupled non-linear differential equations 
from which the TEO can be calculated as a factorized element (that is, as a product of exponentials each containing only one 
generator of the algebra) of the correspondent Lie group. 
It is worth emphasizing that in most cases such solutions must be computed numerically, and the same is true for other exact solutions such as those involving invariant quantities \cite{LEWIS-1969, Shen_2003}, for example. 
A different algebraic approach, based on Baker-Campbell-Haussdorf (BCH)-like relations obtained recently \cite{DMT-BCH-2020}, 
provides a simple recursive way to directly compute the TEO of physical systems described by TD Hamiltonians which are 
written as linear combinations of the generators of the 
$\mathfrak{su}(1,1)$, $\mathfrak{su}(2)$ and $\mathfrak{so}(2,1)$ Lie algebras. 
Notably, its numerical implementation is easy and limited only by computational capacity, and such approach has proven to be 
efficient in the study of the time evolution of a TD-quantum harmonic oscillator \cite{DMT-PS-2020, DMT-JPHYSB-2021} 
and a system of two coupled TD-qubits \cite{Duriez_2022}.  
The main purpose of this work is to develop a formalism to recursively solve the differential equations that arise from the use of the Wei-Norman theory and to use the latter theory to directly obtain effective quantum Hamiltonians for the physical systems described by these differential equations.

In Section {\ref{FromStoRE}} we present the mathematical scenario and apply the Wei-Norman theory to quantum systems of the 
$\mathfrak{su}(1,1)$, $\mathfrak{su}(2)$ and $\mathfrak{so}(2,1)$ Lie algebras, arriving to the time-dependent complex Riccati equation (TDCRE). Moreover, we obtain the complete unitarity criteria for the TEO in the factorized representation. 
In Section {\ref{NewBCH}} we use an explicit solution for time-dependent quantum systems to solve the TDCRE recursively as generalized continued fractions. 
In Section {\ref{BREq}} we apply our results. Initially, we map the so-called Bloch-Riccati equation (BRE)
\cite{Silver_1985, de_Graaf_2019} 
into an effective quantum Hamiltonian of the $\mathfrak{su}(2)$ Lie algebra that can be performed by a TD-qubit. 
Subsequently, we solve the BRE numerically considering a complex hyperbolic secant pulse in the Rabi frequency and with a parameter domain where spin inversion phenomenon is generated. Comparison with the analytical results shows excellent agreement.  
Section {\ref{CONclu}} is left for conclusions and final comments.


\section{From Schr\"odinger to Riccati}\label{FromStoRE}

In this section we initially set the mathematical scenario for the simultaneous treatment of quantum systems 
described by TD hermitian Hamiltonians that are written as linear combinations of the generators of 
the aforementioned algebras. 
Let us consider the following Hamiltonian: 
\begin{equation}
\hat{H}(t)=\eta_{+}(t)\hat{T}_{+}+\eta_{c}(t)\hat{T}_{c}+\eta_{-}(t)\hat{T}_{-} \, ,
\label{eq:TDHSU}
\end{equation}
where the $\eta$-coefficients are in principle arbitrary (at least piecewise constant) scalar functions of time,
and the $\hat{T}$'s are time-independent operators satisfying 
\begin{equation}
\left[\hat{T}_{-},\hat{T}_{+}\right]=2\epsilon\hat{T}_{c} \:\: \mbox{and} \:\: \left[\hat{T}_{c},\hat{T}_{\pm}\right]=\pm\delta\hat{T}_{\pm}\, .
\label{eq:algebraK}
\end{equation}
The parameters $\epsilon$ and $\delta$ allows us to identify the operators $\hat{T}$ 
as the generators of the $\mathfrak{su}(1,1)$, $\mathfrak{su}(2)$ or $\mathfrak{so}(2,1)$ Lie algebras, 
as indicated in the following table: 
\begin{table}[h!]
\centering
\begin{tabular}{ | m{6em} | m{0.8cm}| m{0.4cm} | } 
 \hline
Lie Algebra & $\epsilon$ & $\delta$ \\
 \hline
$\mathfrak{su}(1,1)$  & $1$  & $1$  \\
\hline
 $\mathfrak{su}(2)$  & $-1$  & $1$ \\
\hline
$\mathfrak{so}(2,1)$  & $i/2$  & $i$ \\
\hline
\end{tabular}
\label{table}
		\caption{Relations between the Lie algebras under consideration and parameters $\epsilon$ and $\delta$.}
\end{table}

Let us assume $\hat{T}_{+}=\hat{T}_{-}^{\dagger}$. 
Therefore, from Eqs. (\ref{eq:algebraK}) $\hat{T}_{c}$ is anti-hermitian for the $\mathfrak{so}(2,1)$ algebra 
or hermitian for the other two. 
Using the above, it can be shown that the hermiticity of the Hamiltonian is guaranteed if $\eta_{+}(t)=\eta_{-}^{*}(t)$, with $*$ denoting complex conjugation, while $\eta_{c}(t)$ must be 
either pure imaginary for the $\mathfrak{so}(2,1)$ algebra or real for the other two. 
Accordingly, three independent real-valued functions are needed to define completely the Hamiltonian, 
namely, two for $\eta_{+}(t)$ and one for $\eta_{c}(t)$. 
The Hamiltonian can be thus written as
\begin{equation}
\hat{H}(t)=\eta_{+}(t)\hat{T}_{+}+\eta_{c}(t)\hat{T}_{c}+\eta_{+}^{*}(t)\hat{T}_{-} \, . 
\label{eq:TDHSU2}
\end{equation}
%

The state vector of a quantum system $\left|\psi(t)\right\rangle$ obeys the Schr\"odinger equation 
$i\frac{\partial}{\partial t}\left|\psi(t)\right\rangle=\hat{H}(t)\left|\psi(t)\right\rangle$ ($\hbar = 1$) 
\cite{Sakurai-Book-2014}, and the corresponding TEO is defined by 
$\left|\psi(t)\right\rangle = \hat{U}(t,0)\left|\psi(0)\right\rangle$, where we set the initial time at $t=0$.  
Thereupon, the TEO fulfils the initial condition $ \hat{U}(0,0) = 1\!\! 1$, 
obeys the differential equation
\begin{equation}
i\frac{\partial}{\partial t}\hat{U}(t,0)=\hat{H}(t)\hat{U}(t,0) \, ,
\label{eq:ScroTEO}
\end{equation}
and satisfies the composition property  
\begin{equation}
\hat{U}(t,0) = \hat{U}(t,t_{N-1})\hat{U}(t_{N-1},t_{N-2})\ldots\hat{U}(t_{2},t_{1})\hat{U}(t_{1},0) \, .
\label{eq:CompoTeo}
\end{equation}
There is no general method to find the TEO in Eq. (\ref{eq:ScroTEO}) for an arbitrary TD Hamiltonian. However, 
when symmetries corresponding to the Lie groups can be identified in it, there is a general way to proceed, as we show below.


\subsection{Wei-Norman theory and the Riccati equation}\label{Wei-Norman}

The Wei-Norman theory \cite{Wei_1963, Wei_1964} ensures that, whether a Hamiltonian can be expressed as a 
linear combination of time-independent generators of a finite Lie algebra, 
therefore the TEO can be written as an element of the correspondent Lie group,  
expressed as a product of exponentials of the algebra generators \footnote{A 
recent application of this theory in a system of two coupled TD-quantum 
harmonic oscillators can be found in Ref. [70].}.
Accordingly, for our Hamiltonian in Eq. (\ref{eq:TDHSU2}) we are allowed to consider
the TEO factorized in the following convenient arrangement:
\begin{align}
\hat{U}(t) = e^{\alpha(t)\hat{T}_{+}}e^{\ln(\beta(t))\hat{T}_{c}}e^{\gamma(t)\hat{T}_{-}} \, ,
\label{eq:TEONcomp}
\end{align}
where we suppressed the initial time in our notation. 
Notice that there are $3!$ different but equivalent ways of 
ordering the exponentials of the generators, each arrangement with a different set of coefficients. 
Substituting the above equation together with Eq. (\ref{eq:TDHSU2}) in 
Eq. (\ref{eq:ScroTEO}), 
and with the aid of ordering techniques \cite{Barnett-1997} (similarly to those found in Appendix \ref{appA}), 
we obtain the following set of coupled differential equations for the coefficients of the TEO: 
\begin{align}
& \dot{\alpha} - \delta\frac{\dot{\beta}}{\beta}\alpha + \epsilon\delta\frac{\dot{\gamma}}{\beta^{\delta}}
\alpha^{2}+i \eta_{+}=0 \, , \nonumber\\[3pt]
&\frac{\dot{\beta}}{\beta}-2\epsilon\frac{\dot{\gamma}}{\beta^{\delta}}\alpha +i\eta_{c}=0 \, , \nonumber\\[3pt]
&\frac{\dot{\gamma}}{\beta^{\delta}}+i\eta_{+}^{*}=0 \, ,
\label{eq:sysofdiffeq}
\end{align}
satisfying the initial conditions $\alpha=0$, $\gamma=0$ and $\beta=1$ at $t=0$, and 
where the overdot indicates time derivative. 
Notice that in the latter expressions we have omitted the temporal dependence in the argument of the functions for simplicity of notation. We shall do that along the text whenever there is not risk of confusion. 
The decoupling of the above equations leads to 
\begin{equation}
\dot{\alpha}+\epsilon\delta(i \eta_{+}^{*})\alpha^{2}+ \delta(i \eta_{c})\alpha +i \eta_{+}=0 \, , 
\label{eq:Riccati}
\end{equation}
which is a TD complex Riccati equation (TDCRE) in $\alpha$. Actually, the above equation represents three families of TDCREs, each associated with one of the Lie algebras presented in Table \ref{table}. Note that $\eta_{+}$ is the parameter associated to the non-linearity of the Eq. (\ref{eq:Riccati}), and the solution for $\eta_{+}=0$ with the mentioned initial condition for $\alpha$ is the trivial one, namely, $\alpha(t)=0$. Once the equation for $\alpha$ is solved one can find $\beta$ from Eqs. (\ref{eq:sysofdiffeq}) as  
\begin{equation}
\beta(t)=\exp\left\{-2i\epsilon\int_{0}^{t}\eta_{+}^{*}(t')\alpha(t')dt'-i\int_{0}^{t}\eta_{c}(t')dt' \right\}\, ,
\label{eq:sollambdac}
\end{equation}
and then $\gamma$ can be calculated as 
\begin{equation}
\gamma(t)=-i \int_{0}^{t}\eta_{+}^{*}(t')\beta^{\delta}(t')dt' \,.
\label{eq:solgmabdac}
\end{equation}
We can conclude, therefore, that the solution of the TEO corresponding to the TD Hamiltonian in Eq. (\ref{eq:TDHSU2}) is equivalent to solve the TDCRE of Eq. (\ref{eq:Riccati}). 
Notice that the TDCRE is of great importance, e.g., in mathematics \cite{Ricc-Eq-1991}, physics \cite{CARI_ENA_2000, Markovich_2017, Schuch_2018, Faraoni_2022} and optimal control theory \cite{Zelikin_2000, lewis_2012optimal}. 
More specifically, in quantum physics the Hamiltonian of many prominent systems, such as coupled harmonic oscillators, models for spin and coupled spins, a charged particle moving in a magnetic field, or coupled two-photon lasers, can be put in the form of Eq. (\ref{eq:TDHSU2}) (the above and others examples can be found in Ref. \cite{Shen_2003} and the references therein). 
Recall, the TDCRE has some known analytical solutions but, in general, 
it must be solved numerically \cite{Tsai_2010}. 

One important nontrivial example of a TDCRE with a known analytical solution 
is the so-called Bloch-Riccati equation (BRE) \cite{Silver_1985}, which 
we shall use in the Sec. \ref{BREq} to do a consistency test of our results. 
Hence, it is appropriate to look at the problem from a reverse perspective and ask: When can we use our results if we start with a generic TDCRE as
\begin{equation}
\dot{\alpha} + b_{0}\alpha^{2} + b_{1}\alpha + b_{2}=0 \, ,
\label{eq:GenericRiccati}
\end{equation}
with $b_{0}, b_{1}$ and $b_{2}$ arbitrary complex functions of time?   
A comparison between the above equation and Eq. (\ref{eq:Riccati}) allows us to conclude that, 
to apply our solution, $b_{1}$ must be a pure imaginary function of time, 
$b_{2}$ arbitrary, and $b_{0}$ depending on the algebra as  
$b_{0}=\frac{b_{2}^{*}}{2}$ ($\mathfrak{so}(2,1)$), $b_{0}=-b_{2}^{*}$ ($\mathfrak{su}(1,1)$) or 
$b_{0}=b_{2}^{*}$ ($\mathfrak{su}(2)$). 
Importantly, using the above prescription it is possible to relate TDCREs  directly with quantum Hamiltonians of the mentioned algebras, as we shall show in Sec. \ref{BREq}.   


\subsection{Unitarity criteria}\label{unitary}

As previously mentioned, three independent real-valued functions are needed to fully define the Hamiltonian, and thus the same is true to fully describe the TEO \cite{GILMORE-2012}. 
However, although these functions should be identified directly from the constraints derived from the unitarity criteria for the TEO, we note that the latter is not easily found in the literature for the elements in the representation of Eq.  (\ref{eq:TEONcomp}). 
Therefore, due to the importance of unitarity in physical systems, 
and also as a per se relevant mathematical result, 
in Appendix \ref{appB} we take advantage of the algebraic methods developed in Ref. \cite{DMT-BCH-2020}  to demonstrate the complete unitarity criteria that we list next.
For an arbitrary element of the Lie groups under consideration written as 
$\hat{G}=e^{\left|\alpha\right|e^{i \theta}\hat{T}_{+}}e^{\ln(\left|\beta\right|e^{i \xi})\hat{T}_{c}}e^{\left|\gamma\right|e^{i \phi}\hat{T}_{-}}$, 
the first constraint is 
\begin{equation}
\left|\alpha\right|=\left|\gamma\right|\,,   
\label{inversescondall}
\end{equation}
independently of the group. 
For the $SO(2,1)$ Lie group the  remaining two constraints are 
\begin{equation}
e^{-\xi}=1+\frac{\left|\alpha\right|^{2}}{2}\,\,\,\,\,\,\mbox{and} \,\,\,\,\,\, 
\ln\left|\beta\right|=\theta+\phi \pm n\pi\,, 
\label{unitarityso}
\end{equation}
with $n=1,2,...\,$. On the other hand, for the $SU(1,1)$ and $SU(2)$ Lie groups the remaining two constraints are
\begin{equation}
\left|\beta\right|+\epsilon\left|\alpha\right|^{2}=1\,\,\,\,\,\,\mbox{and} 
\,\,\,\,\,\, \xi=\theta +\phi \pm n\pi\, ,
\label{unitaritysu}
\end{equation}
%
with $n=1,2,...\,$. To the best of the author's knowledge, this is the first time that the above relations are formally 
calculated for the groups under consideration. 
Nevertheless, in an important paper of Truax \cite{truax-1985} he showed that, starting with an unfactorized representation for an unitary element of the $SU(1,1)$ and $SU(2)$ Lie groups, the factorized representation remains unitary.


\section{Time evolution of quantum systems}\label{NewBCH}

We now follow a simple recursive solution recently developed in 
Ref. \cite{DMT-BCH-2020} to calculate the TEO corresponding to the Hamiltonian in Eq. (\ref{eq:TDHSU}).
There, the authors considered a time-splitting in $N$ intervals of equally small enough size $\tau=t/N$, 
such that the Hamiltonian coefficients, and therefore the Hamiltonian itself, can be regarded as constant 
in each $j$-th time-interval ($j=1,2,...,N$). Formally, this implies that the present solution will coincide 
with the exact one only in the limit $N\rightarrow\infty$ (and $\tau\rightarrow 0$). 
Nevertheless, for numerical implementation of this method, it is enough to choose $\tau$ to be much
smaller than the typical timescale of the Hamiltonian coefficients. 
For our Hamiltonian in Eq. (\ref{eq:TDHSU2}) let us write these functions compactly, 
henceforth, as $\boldsymbol{\eta}=(\eta_{+},\eta_{c},\eta_{+}^{*}) \,$. 
Without loss of generality we define their $j$-th value as $\boldsymbol{\eta}_j\equiv\boldsymbol{\eta}(t=j\tau)$, 
where $\boldsymbol{\eta}_j=(\eta_{j+},\eta_{jc},\eta_{j+}^{*})$, and the 
correspondent $j$-th TEO can be thus written as a Lie group element in the \textit{unfactorized} representation 
$\hat{U}_{j}=\exp{\lambda_{j+}\hat{T}_{+}+ \lambda_{jc}\hat{T}_{c}+\lambda_{j-}\hat{T}_{-}}$, 
where $\boldsymbol{\lambda}_j\equiv(\lambda_{j+},\lambda_{jc},\lambda_{j-})=-i\tau\boldsymbol{\eta}_j$. 
Using BCH-like relations (see Appendix \ref{appA} for details) each $\hat{U}_{j}$ can be re-expressed in the 
factorized representation $\hat{U}_{j}=e^{\Lambda_{j+}\hat{T}_{+}}e^{\ln(\Lambda_{jc})\hat{T}_{c}}e^{\Lambda_{j-}\hat{T}_{-}}$,  
with the coefficients given by 
\begin{align}
\label{truej4}
\Lambda_{jc}=&\left(\cosh(\nu_{j})-\frac{\delta\lambda_{jc}}{2\nu_{j}} \sinh(\nu_{j})\right)^{-\frac{2}{\delta}},\\
\label{truej5}
\Lambda_{j\pm}=&\frac{2\lambda_{j\pm} \sinh(\nu_{j})}{2\nu_{j} \cosh(\nu_{j})-\delta\lambda_{jc}\sinh(\nu_{j})} \, ,
\end{align}
and 
\begin{equation}
\nu_{j}^{2} = \left(\frac{\delta\lambda_{jc}}{2}\right)^{2}-\delta\epsilon\lambda_{j+}\lambda_{j-} \, .
\label{eq:nu}
\end{equation}
Since each $\hat{U}_{j}$ is a Lie group element and the TEO is given by the composition of them, 
then the total TEO must also be an element of the Lie group, so that it can be written in the form 
\begin{equation}
\hat{U}(t,0) = e^{\alpha_{N}\hat{T}_{+}}e^{\ln(\beta_{N})\hat{T}_{c}}e^{\gamma_{N}\hat{T}_{-}} \, ,
\label{eq:TEONcompRicc}
\end{equation}
where the choice of the coefficients in the previous expression is not coincidental. 
Indeed, comparing it with Eq. (\ref{eq:TEONcomp}) it is clear that 
the solution for $\alpha$, $\beta$ and $\gamma$ is the same. 
Using the composition rule shown in Appendix \ref{appB}, the coefficients in the above equation can be written recursively as 
\cite{DMT-BCH-2020}
\begin{align}
\alpha_{j}&=\,\Lambda_{j+}+\frac{\alpha_{(j-1)}(\Lambda_{jc})^{\delta}}{1-\epsilon\delta\alpha_{(j-1)}\Lambda_{j-}} \, , \label{eq:alpharN} \\
\beta_{j}&=\,\frac{\beta_{(j-1)}\Lambda_{jc}}{\left(1-\epsilon\delta\alpha_{(j-1)}\Lambda_{j-}\right)^{\frac{2}{\delta}}} \, , \label{eq:betaN} \\
\gamma_{j}&=\,\gamma_{(j-1)}+\frac{\Lambda_{j-}(\beta_{(j-1)})^{\delta}}{1-\epsilon\delta\alpha_{(j-1)}\Lambda_{j-}} \, , \label{eq:gammaN}
\end{align}
with $\alpha_{1}=\Lambda_{1+}$, $\beta_{1}=\Lambda_{1c}$, $\gamma_{1}=\Lambda_{1-}$ and $j=1,2,...,N$. 
Notice that the $\alpha$ parameter is an independent 
term of $\beta$ and $\gamma$, since the later 
two need the former to be calculated. Furthermore, 
it can be written as
\small
\begin{equation}
\alpha_{j}=\Lambda_{j+}-\cfrac{(\Lambda_{jc})^{\delta}}{\epsilon\delta\Lambda_{j-}-\cfrac{1}{\Lambda_{(j-1)+}-
\cfrac{(\Lambda_{(j-1)c})^{\delta}}
{\epsilon\delta\Lambda_{(j-1)-} \, -\cfrac{1}{\ldots -\cfrac{1}{\Lambda_{1+}}}}}} \, , \\
\label{eq:gammarecursive}
\end{equation}
\normalsize
\textit{i.e.}, as a generalized continued fraction (GCF). 
This kind of mathematical objects are important in the realm of complex analysis and 
are specially useful to study analyticity of functions as well as number theory, 
among other fields (see Ref. \cite{Baumann_2019} and references therein). 
More importantly, its numerical implementation is straightforward and 
limited only by computer capacity, as demonstrated in Refs. \cite{DMT-PS-2020, DMT-JPHYSB-2021, Duriez_2022}. 

Note that, although $\alpha$ is written recursively, the analytical calculation of its derivatives 
can be done by finding the differential equation that it satisfies, which we have demonstrated in 
the Sec. \ref{FromStoRE} to be the TDCRE. Furthermore, there is a way to demonstrate this last result directly from the recursive solution for $\alpha$ given in Eq. (\ref{eq:alpharN}).  
To do this, let us consider the limit of small time intervals satisfying $\abs{\eta_j}\tau\ll 1\,$, so that $\nu_j\ll 1\,$. Up to first order in $\tau\,$, we have $\sinh(\nu_j)\approx\nu_j$ and $\cosh(\nu_j)\approx 1\,$. In this case, one trivially finds $\left(\Lambda_{jc}\right)^{\delta} \approx 1-i\delta\,\eta_{jc}\tau\,$, $\Lambda_{j+}\approx  -i\eta_{j+}\tau\,$, and $\Lambda_{j-}\approx -i\eta^{*}_{j+}\tau\,$. Therefore $(1-\epsilon\delta\Lambda_{j-}\alpha_{j-1})^{-1} \approx 1 - i\epsilon\delta\,\eta^{*}_{j+}\,\alpha_{j-1}\,\tau\,$, and up to first order in $\tau\,$ the recurrence relation for $\alpha$ becomes
	\begin{equation}
		\alpha_j \approx \alpha_{j-1} - \tau\left(i\epsilon\delta\,\eta^{*}_{j+}\,\alpha^2_{j-1} + i\delta\eta_{jc}\alpha_{j-1} + i\eta_{j+}\right)
		\,.
	\end{equation}
	When $\tau\to 0\,$, the finite difference ratio in $\alpha$ approaches a derivative resulting in the Riccati equation (\ref{eq:Riccati}). In this way, the connection between the generalized continued fraction and the complex Riccati equation for $\alpha$ is straightforward. 


As a particular case, we can realize that $j=1$ corresponds to the exact analytic solution 
for a sudden change (also called a jump or a quench) in the Hamiltonian coefficients at $t=0$. 
Moreover, Eqs. (\ref{eq:alpharN})-(\ref{eq:gammaN}) represent the exact analytic solution for a sequence of N jumps 
equally spaced in time of the Hamiltonian coefficients. 
For one jump at $t=0$, they can be written as 
$\boldsymbol{\eta}(t) = \boldsymbol{\eta}^{o} + (\boldsymbol{\eta}^{f} - \boldsymbol{\eta}^{o})\Theta(t)\, ,$
where $\boldsymbol{\eta}^{o}\equiv\boldsymbol{\eta}(t<0)\,$, $\boldsymbol{\eta}^{f}\equiv\boldsymbol{\eta}(t\geq0)\,$, 
and $\Theta$ is the usual Heaviside step function. The corresponding TEO is therefore given by 
$\hat{U}_{1}(t,0) = e^{\Lambda_{1+}\hat{T}_{+}}e^{\ln(\Lambda_{1c})\hat{T}_{c}}e^{\Lambda_{1-}\hat{T}_{-}} \,$, 
with the $\Lambda$-functions given by Eqs. (\ref{truej4}) and (\ref{truej5}) evaluated for 
$\boldsymbol{\lambda}_1=-i t(\eta_{+}^{f},\eta_{c}^{f},\eta_{+}^{*f})$. 
Notice that, as we are using essentially the composition rule for the elements of the groups corresponding to the algebras under consideration, 
our solution also contemplates the calculation of the TEO for systems with a finite number of jumps in their parameters 
that are not necessarily equally spaced in time e.g., the case of a quantum harmonic oscillator with frequency jumps 
\cite{JANSZKY-1992, JANSZKY-TE-1994, DMT-BJP-2019, Stanley_2022}, 
being useful in the construction of squeezed states of atomic motion in optical lattices \cite{Xin_2021}. 
Moreover, they can also be used to calculate the arbitrary composition of squeeze operators, rotation operators and many other interesting 
unitary operators of the Lie algebras under consideration (see, for instance, Refs. \cite{Benjamin_2022nano, Bhattacharjee_2022}). 

To end this section it is important to note that the constraints obtained from the unitarity in 
Eqs. (\ref{inversescondall}), (\ref{unitarityso}) and (\ref{unitaritysu}) can be used as a fundamental test 
for the numerical implementations of our results, once the $\Lambda$-functions in Eqs. (\ref{truej4}) and (\ref{truej5}) 
must satisfy them at any time (for any $j$), and the same is true for Eqs. (\ref{eq:alpharN}), (\ref{eq:betaN}) and (\ref{eq:gammaN}).


\section{Effective Hamiltonian and solution of the Bloch equations}\label{BREq}
 
Since its publication in 1946 \cite{Bloch_1946}, the Bloch equations became of fundamental importance in the realm of nuclear magnetic resonance (NMR). They provide a quantitative description of any NMR experiment that involves radio frequency pulses, which are at the heart of all modern NMR experiments \cite{de_Graaf_2019}. 
In this section we shall use our results to identify the effective quantum Hamiltonian for the Bloch-Riccati equation and numerically recover the remarkable results of Silver et al \cite{Silver_1985} as an application of our results.
Following this reference, 
the Bloch equations in the rotating frame, neglecting relaxation terms, can be written as 
\begin{align}
& \dot{M} + i \Delta\omega M + M_{z}\Omega(t)=0  \, , \nonumber\\[3pt]
& \dot{M}_{z} - \frac{i}{2} \left(M\Omega^{*}(t)-M^{*}\Omega(t)\right)=0 \, ,
\label{eq:Blocheq}
\end{align}
where $M$ is the complex magnetization in the $x-y$ plane, $M_{z}$ the longitudinal magnetization, and 
$\Omega(t)=-g(B_{1x}+iB_{1y})$ is the complex TD driving function. 
Using the following definition 
\begin{equation}
f=\frac{M}{M_{o}+M_{z}} \, ,
\label{eq:fiufiu}
\end{equation}
where $M_{o}$ is the equilibrium magnetization, Eqs. (\ref{eq:Blocheq}) can be transformed 
into the Bloch-Riccati equation (BRE) \cite{Silver_1985}, namely,
\begin{equation}
\dot{f}-\frac{i}{2}\Omega^{*}(t) f^{2}+ i\Delta\omega f +\frac{i}{2}\Omega(t) =0\, . 
\label{eq:BlochRiccati}
\end{equation}
Notice that the above equation is identical to Eq. (\ref{eq:Riccati}) for the particular case of the 
Lie algebra $\mathfrak{su}(2)$ ($\delta=1$ and $\epsilon=-1$), 
with the identifications $\eta_{+}=\Omega(t)/2$ and $\eta_{c}=\Delta\omega$. Therefore, using our results of section \ref{Wei-Norman} the corresponding effective Hamiltonian is simply given by
\begin{equation}
\hat{H}(t)=\frac{1}{2}\Omega(t)\hat{T}_{+}+\Delta\omega\hat{T}_{c}+\frac{1}{2}\Omega^{*}(t)\hat{T}_{-} \, .
\label{eq:TDHBlochRicc}
\end{equation}
Furthermore, if we consider a realization of this algebra for the Pauli operators $\sigma_{+}$, $\sigma_{-}$ and $\sigma_{z}$, 
with $\sigma_{\pm}=\sigma_{x}\pm i\sigma_{y}$, we note that the above Hamiltonian will be exactly the same of a 
TD-qubit \cite{de-Clercq-2016, DMT-BCH-2020} with $\Omega(t)$ the Rabi frequency and 
$\Delta\omega$ an effective detuning. 
We conclude, therefore, that the time evolution of a TD-qubit is, among several other possible TD systems of the 
$\mathfrak{su}(2)$ Lie algebra, a set up equivalent to RMN described by the Bloch equations in the rotation frame and 
with relaxation terms neglected, a result that is in agreement with the well-known connection between quantum optics 
and RMN \cite{de_Graaf_2019}. Note that, although there are some exact solutions for a TD-qubit with 
certain driving terms (see e.g., Ref. \cite{Enriquez_2018}), our general approach allows us to consider  
arbitrary complex TD functions for the Rabi frequency as well as arbitrary real-valued functions for the detuning. 

\begin{figure}[h!]
\includegraphics*[width=9.0cm]{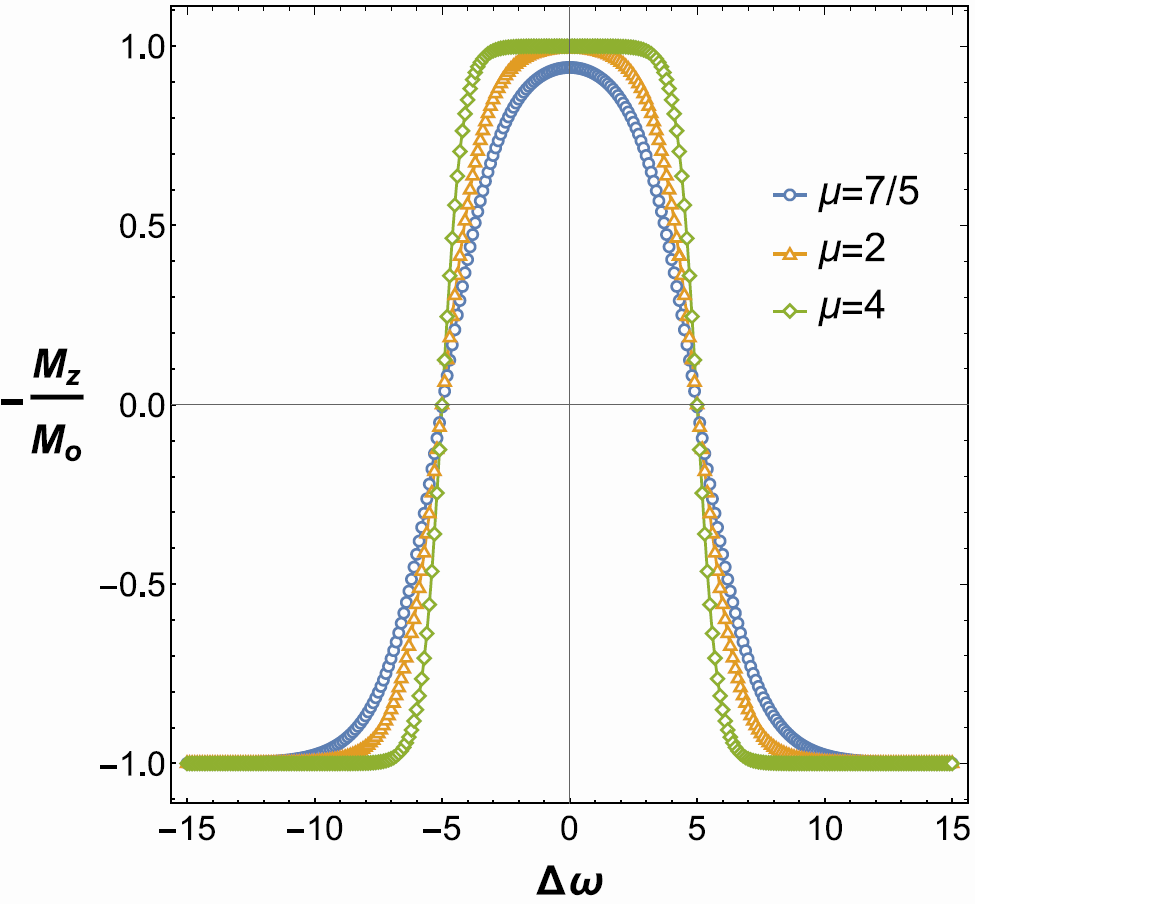}
\caption{Spin inversion as a function of the detuning using the driving in Eq. (\ref{eq:SechPulse}) 
for different values of the real parameter $\mu$. 
Solid lines are calculated with the analytic expression of Eq. (\ref{eq:inversion}), while the points are numerically calculated with 
our general approach.}
\label{fig:Consistency}
\end{figure}

A nontrivial emblematic example with analytical solution for the BRE and with relevant experimental applications 
in NMR comes from the use of a complex hyperbolic secant pulse as the driving function for the Rabi frequency \cite{Silver_1985}. 
Moreover, choosing appropriately the domain of the involved functions, spin inversion phenomenon can be achieved.  
As a consistency test, we now shall recover the results for such driving. 
Let us consider the following family of functions for the Rabi frequency:
\begin{equation}
\Omega(t)=\Omega_{o}\left(\sech{\chi (t-t_{o})}\right)^{1+i\mu} \, , 
\label{eq:SechPulse}
\end{equation}
where $\mu$ is a real constant and $\Omega_{o}$ is the pulse amplitude. 
Using the above driving, the BRE (\ref{eq:BlochRiccati}) can be transformed in a hypergeometric equation 
with known solutions. After taking into account the initial conditions, the authors in Ref. \cite{Silver_1985} 
found for the stationary solution of the magnitude of $f(t)$  the following simple expression 
\begin{equation}
\left|f\right|^{2}_{t\rightarrow\infty}= \frac{\cosh^{2}{(\pi\mu/2)}-\cos^{2}{(\pi y)}}{\cosh^{2}{(\pi\Delta\omega/2\chi)}-\sin^{2}{(\pi y)}} \, ,
\label{eq:finalargalpha}
\end{equation}
with
\begin{equation}
y=\left\{\left(\frac{\Omega_o}{2\chi}\right)^{2}-\left(\frac{\mu}{2}\right)^{2}\right\}^{1/2}\, , 
\label{eq:finalargalpha1}
\end{equation}
and where they considered solutions for $y$ real and $2 y \neq1,2,3,...$. 
The quantity that serves to predict the spin inversion is given by 
$\frac{M_z}{M_o}=\frac{1-\left|f\right|^{2}_{t\rightarrow\infty}}{1+\left|f\right|^{2}_{t\rightarrow\infty}}\,$. 
Using Eq. (\ref{eq:finalargalpha}), it can be shown that
\begin{equation}
\frac{M_z}{M_o}=\tanh{\varphi_1}\tanh{\varphi_2}+\sech{\varphi_1}\sech{\varphi_2}\cos{\varphi_3}\, .
\label{eq:inversion}
\end{equation}
where 
\begin{align}
\label{argumentbres}
\varphi_1=&\pi\left\{\frac{\Delta\omega}{2\chi}+\frac{\mu}{2}\right\} \, , \,\,\,\,
\varphi_2=\pi\left\{\frac{\Delta\omega}{2\chi}-\frac{\mu}{2}\right\} \nonumber\\[4pt]
&\mbox{ and }\,\,\,\, \varphi_3=\small\pi\left\{\left(\frac{\Omega_o}{\chi}\right)^{2}-\mu^{2}\right\}^{1/2} \, .
\end{align}\normalsize
Spin inversion is achieved if the above quantity changes from $\frac{M_z}{M_o}=1$ to $\frac{M_z}{M_o}=-1$ 
(and vice versa). Moreover, notice from Eq. (\ref{eq:inversion}) that without phase modulation 
($\mu=0$) and whenever that $\Omega_o=2n\chi$ with 
$n=1,2,...$ there is an excursion of the magnetization ($M_z=M_o$). 
On the other hand, in the limit of $\mu\rightarrow\infty$, $\chi\rightarrow 0$, with 
$\mu\chi\rightarrow C$ a constant, and $\Omega_o\geq C$, the  magnetization is inverted over all frequencies ($M_z=-M_o$).  
To capture this phenomenon and taking into account the considered analytical solutions, we chose for the numerical 
calculations the following values of the parameters in arbitrary units: 
$\Omega_o=10$, $\chi=\Omega_o/2\mu$ and $\mu=\left\{1.4,2,4\right\}$ for the phase modulation parameter, with fixed values of $\Delta\omega$ 
sweeping the interval $-15\leq\Delta\omega\leq15$. We also chose a time interval for the analysis of $t \in [0,40]$. 
In Fig. \ref{fig:Consistency}, we plot in solid lines the analytic solution of Eq. (\ref{eq:inversion}) 
for the chosen values of the parameters. 
Recall that the intermediary numerical calculations of Eqs. (\ref{truej4})-(\ref{eq:nu}) 
allow for the testing of the code using the unitary relations given in 
Eqs. (\ref{inversescondall}) and (\ref{unitaritysu}) ($\epsilon=-1$). 
Now, considering the 
time-splitting in $N=8\times10^{3}$ intervals for each value (point) of $\Delta\omega$, we numerically calculate a total of 300 points 
within the effective detuning interval to evaluate the same quantity with our formalism. 
The calculations took less than 20 minutes in a laptop machine and are plotted as the dot patterns in Fig. \ref{fig:Consistency}. 
As it can be noted, the matching is excellent, allowing us to validate our results. 
Recall that our formalism allows one to fully calculate the TEO corresponding to the RMN effective 
Hamiltonian obtained in Eq. (\ref{eq:TDHBlochRicc}) \textit{i.e.}, the $\beta$ and $\gamma$ functions in addition to $\alpha$, 
and, consequently, to evolve any initial state. Nevertheless, the only necessary parameter to describe spin inversion in this case is $\alpha$. 
This is because the Bloch equations (\ref{eq:Blocheq}) consider intrinsically the initial state of the system as the ground state, 
where $\alpha$ is the protagonist of the dynamics. 
Therefore, we can expect that in RMN analysis using arbitrary drivings and initial states, our results will be useful.



\section{Conclusions}\label{CONclu}

In the first part of this work we derived a TD complex Riccati equation (TDCRE) from the application of
the Wei-Norman theory in TD systems of the $\mathfrak{su}(1,1)$, $\mathfrak{su}(2)$ and $\mathfrak{so}(2,1)$ Lie algebras, 
and so we obtained all the necessary and sufficient conditions for the unitarity of the elements of the correspondent given Lie groups in the factorized representation.  
This result is important as unitarity guarantees probability conservation, and to the best of the author's knowledge, this is the first time these complete conditions are listed.  
Then, in the second part of this work we used a solution for time-dependent quantum systems to solve the TDCRE 
recursively as generalized continued fractions (GCF), which are optimal for numerical implementations. 
The formalism we developed also allows us to associate effective Hamiltonians directly to TDCREs,   
as we showed in the third part of this work for the so-called Bloch-Riccati equation (BRE), 
mapping it in an effective quantum Hamiltonian of the $\mathfrak{su}(2)$ Lie algebra. 
Then, as an application, but also as a consistency test, we numerically calculated the solution of 
the BRE for a complex hyperbolic secant pulse generating spin inversion and compare it to analytical results, showing excellent agreement. 
Our results are quite general and can be used not just to solve the TEO of any TD system of the algebras at issue and its related 
TDCRE, but also TDCREs  
that do not need to be related to quantum systems. 
For instance, Newton’s laws can be put in the form of Riccati equations under certain conditions \cite{Nowakowski_2002}, 
and therefore, it should be possible to 
apply the methods discussed in this work.

Our results can be straightforwardly extended for other non-linear differential equations that can be derived 
from the TDCRE, e.g., the (dissipative) Ermakov equation \cite{Schuch_2008, Schuch_2014, Blanco_Garcia_2018} with its 
respective invariant, paving the way for new possibilities in the quantum-classical connection \cite{Schuch_2018}. 
Furthermore, our results can be also extended for other algebras with a higher number of generators \cite{Raffa_2022}, 
arriving to different sets of differential equations that could be solved in terms of GCFs, also 
amplifying the possibilities of investigating more complex systems as coupled TD-quantum harmonic oscillators 
\cite{Bruschi_2013, urzua-2-2019, Bruschi_2020, Bruschi_2021} or coupled TD-qubits \cite{Rau_2005, Duriez_2022}. 
Our results related to GCFs could be also useful in the analysis of quantum systems using differential Galois theory \cite{Acosta_Hum_nez_2013}, in extensions for non-hermitian time-dependent systems \cite{Zelaya_2021}, in the search for special symmetries leading to analytical solutions \cite{Gomez_2010}, 
as well as in the analysis of the Lie-Scheffers theorem \cite{CARI_ENA_1998, Cari_ena_2015}. 
Finally, we can expect applications of our results in several branches of physics, as they encompasses 
the solution of the TEO of important fundamental systems as TD-qubits and TD-quantum harmonic oscillators, 
being useful in the control of quantum systems \cite{Ahlbrandt_1996, Li_2009}, in the design of shortcuts to adiabaticity \cite{guery-2019}, 
in the harnessing of nonadiabatic excitations promoted by quantum critical points \cite{Abah_2022}, 
in the description of ion traps dynamics \cite{Mihalcea_2022}, 
in quantum thermodynamics \cite{Kosloff_2017,Elouard_2020, Dann_2020, Fei_2022}, 
in the study of (cavity) optomechanical systems \cite{Bruschi_2018, Qvarfort_2019, Bruschi_2019, Qvarfort_2021} 
or in quantum interference of levitated nano rotors \cite{Benjamin_2022nano},
among many others.

\begin{acknowledgments}


D.M.T. thanks J. C. Correa, R. C. L. Bruni and L. Pires for enlightening discussions.
The authors thank the Brazilian agencies for scientific and technological research, namely,  
the Coordination for the Improvement of Higher Education Personnel - CAPES, 
the National Council for Scientific and Technological Development - CNPq, 310365/2018-0 (C.F.) 
and Fundação de Amparo à Pesquisa do Estado do Rio de Janeiro - FAPERJ, 2021042663 (D.M.T.), 
for partial financial support.

\end{acknowledgments}


\bibliographystyle{unsrt}

\bibliography{Biblio}


\appendix


\section{BCH-like relations and unitary criteria} \label{appB}

The BCH-like relations developed in Ref. \cite{DMT-BCH-2020} 
are, essentially, the composition rule for the elements of the groups corresponding to the algebras under consideration, 
once they are written in the factorized representation of Eq. (\ref{eq:TEONcomp}). More specifically, given two arbitrary elements 
$\hat{G}_{1}=e^{\tilde{\alpha}\hat{T}_{+}}e^{\ln(\tilde{\beta})\hat{T}_{c}}e^{\tilde{\gamma}\hat{T}_{-}}$ 
and $\hat{G}_{2}=e^{\alpha\hat{T}_{+}}e^{\ln(\beta)\hat{T}_{c}}e^{\gamma\hat{T}_{-}}$, 
their product is another element of the group, namely, 
\begin{equation}
\hat{G}=\hat{G}_{2}\hat{G}_{1}=e^{\zeta_{+}\hat{T}_{+}}e^{\ln(\zeta_{c})\hat{T}_{c}}e^{\zeta_{-}\hat{T}_{-}} \, ,
\label{eq:2compgamalta}
\end{equation}
where 
\begin{align}
\label{inverse}
\zeta_{+}=&\alpha+\frac{\tilde{\alpha}\beta^{\delta}}{1-\epsilon\delta\tilde{\alpha}\gamma} \, , \,\,\,\,
\zeta_{c}=\frac{\tilde{\beta}\beta}{\left(1-\epsilon\delta\tilde{\alpha}\gamma\right)^{\frac{2}{\delta}}} \nonumber\\[4pt]
&\mbox{ and }\,\,\,\, \zeta_{-}=\tilde{\gamma}+\frac{\gamma(\tilde{\beta})^{\delta}}{1-\epsilon\delta\tilde{\alpha}\gamma} \, .
\end{align}
Notice, the inverse product \textit{i.e.}, $\hat{G}_{1}\hat{G}_{2}$, leads to identical relations as in Eqs. (\ref{inverse}) but interchanging the 
letters having tilde with those that do not. Suppose that we know $\hat{G}_{2}$ and we desire to obtain its inverse, namely $\hat{G}_{1}$. 
Accordingly, their product must equals the identity, \textit{i.e.}, $\hat{G}_{2}\hat{G}_{1}=\hat{G}_{1}\hat{G}_{2}= 1\!\! 1$. 
In Eq. (\ref{inverse}) this implies $\zeta_{+}=0$, $\zeta_{c}=1$ and $\zeta_{-}=0$, from which we obtain 
$\tilde{\alpha}=  -\frac{\alpha}{l}\,$, $\tilde{\beta}= \frac{\beta}{l^{\frac{2}{\delta}}} \,$ and $\tilde{\gamma}=-\frac{\gamma}{l}$, 
where we defined $l\equiv\beta^{\delta}-\epsilon\delta\,\alpha\,\gamma$. 
It can also be proven that for the inverse product the above relations holds true, and therefore $\hat{G}_{2}$ is the inverse of $\hat{G}_{1}$. 
Let us now define 
\begin{equation}
\alpha=\left|\alpha\right|e^{i \theta}\, , \,\,\,\,\,\,\beta=\left|\beta\right|e^{i \xi}
\,\, \,\,\,\,\,\,\mbox{and} \,\,\,\,\,\,\gamma=\left|\gamma\right|e^{i \phi}.
\label{modulusandarg}
\end{equation}
The unitary condition demands $\hat{G}^{-1}=\hat{G}^{\dagger}$, where $\hat{G}^{\dagger}=
e^{\gamma^{*}\hat{T}_{+}}e^{\ln(\beta^{*})\hat{T}_{c}^{\dagger}}e^{\alpha^{*}\hat{T}_{-}}\,$, 
leading to 
\begin{equation}
\gamma^{*}=-\frac{\alpha}{l}\, , \,\,\,\,\,\,\ln(\beta^{*})\hat{T}_{c}^{\dagger}=
\ln(\frac{\beta}{l^{\frac{2}{\delta}}})\hat{T}_{c}\,\, \,\,\,\,\,\,\mbox{and} \,\,\,\,\,\,\alpha^{*}=-\frac{\gamma}{l}\,.
\label{inversesBCH}
\end{equation}
From the left and right hand side equations above and Eqs. (\ref{modulusandarg}) we obtain the 
following results valid for all the algebras at issue. First, the constraint 
\begin{equation}
\left|\alpha\right|=\left|\gamma\right|\,. 
\label{inversescondall1}
\end{equation}
Second, that $l$ is just a phase once $\left|l\right|=1$. And third, 
that $l^{2}=e^{2i\left(\theta + \phi\right)}$. Recall that, by construction, $\hat{T}_{c}$ 
is anti-hermitian for the $\mathfrak{so}(2,1)$ algebra and hermitian for the other two. 
Accordingly, for the $\mathfrak{so}(2,1)$ algebra, the middle equation in Eqs. (\ref{inversesBCH}) 
implies that $\left|\beta\right|^{2i}=l^{2}$, with $l=\beta^{i}+\frac{\alpha\gamma}{2}$ (see Table I).  
Using the above results together with Eqs. (\ref{modulusandarg}) and (\ref{inversesBCH}), it
is straightforward to show that
\begin{equation}
e^{-\xi}=1+\frac{\left|\alpha\right|^{2}}{2}\,\,\,\,\,\,\mbox{and} \,\,\,\,\,\, 
\ln\left|\beta\right|=\theta+\phi \pm n\pi\,, 
\label{unitarityso1}
\end{equation}
with $n=1,2,...\,$. On the other hand, for the $\mathfrak{su}(1,1)$ and $\mathfrak{su}(2)$ algebras the middle 
equation in Eqs. (\ref{inversesBCH}) implies that $\frac{\beta}{\beta^{*}}=l^{2}$ with $l=\beta-\epsilon\alpha\gamma$, 
leading to 
\begin{equation}
\left|\beta\right|+\epsilon\left|\alpha\right|^{2}=1\,\,\,\,\,\,\mbox{and} 
\,\,\,\,\,\, \xi=\theta +\phi \pm n\pi\, ,
\label{unitaritysu1}
\end{equation}
%
with $n=1,2,...\,$. Using the above results it can be shown that, for all the algebras under consideration  
$l=-e^{i\left(\theta + \phi\right)}$. Finally, notice that the arbitrary composition of squeeze operators, rotation operators and other 
interesting unitary operators of the groups under consideration, can be calculated using these BCH-like relations.

\section{Factorizing group elements} \label{appA}

In this appendix show that an arbitrary element of the Lie groups at issue and given in the 
unfactorized representation, namely, 
\begin{equation}
\hat{G}(\boldsymbol{\lambda})=\exp{\lambda_{+}\hat{T}_{+}+ \lambda_{c}\hat{T}_{c}+\lambda_{-}\hat{T}_{-}} ,
\label{eq:unfac}
\end{equation}
can be factorized in the usual order, as indicated in Eqs. (\ref{truej4}) to (\ref{eq:nu}), or as
\begin{equation}
 \hat{G}(\boldsymbol{\Sigma})= e^{\Sigma_{-}\hat{T}_{-}}e^{\ln(\Sigma_{c})\hat{T}_{c}}e^{\Sigma_{+}\hat{T}_{+}} \, , 
\label{eq:factorization1}
\end{equation}
where
\begin{align}
\label{truesig4}
&\Sigma_{c}=\left(\cosh(\nu)+\frac{\delta\lambda_{c}}{2\nu} \sinh(\nu)\right)^{\frac{2}{\delta}},\\
\label{truesig5}
&\Sigma_{\pm}=\frac{2\lambda_{\pm} \sinh(\nu)}{2\nu \cosh(\nu)+\delta\lambda_{c}\sinh(\nu)} \, ,
\end{align}
with
\begin{equation}
\nu^{2} = \left(\frac{\delta\lambda_{c}}{2}\right)^{2}-\epsilon\delta\lambda_{+}\lambda_{-} \, . 
\label{eq:nu2}
\end{equation}
Firstly, let us re-define Eq. (\ref{eq:unfac}) as the special case $\rho=1$ of the operator
\begin{equation}
\hat{F}_{1}(\rho)=e^{\rho\left(\lambda_{+}\hat{T}_{+}+ \lambda_{c}\hat{T}_{c}+\lambda_{-}\hat{T}_{-}\right)}\, .
\label{fdt}
\end{equation}
The basic idea is to find an equivalent expression in the form
\begin{equation}
\hat{F}_{1}(\rho)= e^{\Sigma_{-}(\rho)\hat{T}_{-}}e^{\ln(\Sigma_{c}(\rho))\hat{T}_{c}}e^{\Sigma_{+}(\rho)\hat{T}_{+}} \, ,
\label{fdt2}
\end{equation}
and therefore, to write the functions $\Sigma_+$, $\Sigma_-$ and $\Sigma_c$ in terms of the small lambdas so that the last two equations are equal. The relations between these two set of functions 
($\boldsymbol{\Sigma}$ and $\boldsymbol{\lambda}$) are known 
as BCH-like relations \cite{Barnett-1997}. Accordingly, we derive both expressions with respect to $\rho$ and 
impose the derivatives to be equal. The derivative of Eq. (\ref{fdt}) is direct, and given by
\begin{equation}
\hat{F}'_{1}= \left(\lambda_{+}\hat{T}_{+}+ \lambda_{c}\hat{T}_{c}+\lambda_{-}\hat{T}_{-}\right) \hat{F}_{1}\, ,
\label{fdt3}
\end{equation}
where the prime indicates derivative with respect to $\rho$. 
On the other hand, the derivative of Eq. (\ref{fdt2}) can be written as
\begin{equation}
\hat{F}'_{1}= \left(\Sigma_{-}\hat{T}_{-}+\frac{\Sigma_{c}'}{\Sigma_{c}}\hat{I}_{1} + \Sigma_{+}\hat{I}_{2}\right) \hat{F}_{1} \, ,
\label{fdt4}
\end{equation}
with
\begin{align}
&\hat{I}_{1}= e^{\Sigma_{-}\hat{T}_{-}}\hat{T}_{c}\, e^{-\Sigma_{-}\hat{T}_{-}}  \nonumber\\[3pt]
&\hat{I}_{2}=	e^{\Sigma_{-}\hat{T}_{-}}\,e^{\ln(\Sigma_{c})\hat{T}_{c}}\hat{T}_{+}\,e^{-\ln(\Sigma_{c})\hat{T}_{c}}\, e^{-\Sigma_{-}\hat{T}_{-}}	\, .
\label{I1I2}
\end{align}
Using the following BCH relation \cite{Barnett-1997}
\begin{align}
e^{\hat{A}}\hat{B}\, e^{-\hat{A}} =& \hat{B}+\left[\hat{A},\hat{B}\right]+\frac{1}{2!}\left[\hat{A},\left[\hat{A},\hat{B}\right]\right]+\nonumber\\[3pt]
																	 & +\frac{1}{3!}\left[\hat{A},\left[\hat{A},\left[\hat{A},\hat{B}\right]\right]\right]+...\, ,
\label{BCHkey}
\end{align}
together with the commutation relations given in Eq. (\ref{eq:algebraK}) we can solve Eqs. (\ref{I1I2}), and then equalize 
Eqs. (\ref{fdt3}) and (\ref{fdt4}) to obtain the following set of coupled differential equations:
\begin{eqnarray}
\label{eqd1}
\Sigma_{-}'+\delta\Sigma_{-}\frac{\Sigma_{c}'}{\Sigma_{c}}+\epsilon\delta(\Sigma_{-})^{2}(\Sigma_{c})^{\delta}\Sigma_{+}'&=&\lambda_{-} \, ,\\
\label{eqd2}
\frac{\Sigma_{c}'}{\Sigma_{c}}+2\epsilon\Sigma_{-}(\Sigma_{c})^{\delta}\Sigma_{+}'&=&\lambda_{c} \, , \\
\label{eqd3}
(\Sigma_{c})^{\delta}\Sigma_{+}'&=&\lambda_{+}\, \, .
\end{eqnarray}
Substitution of Eq. (\ref{eqd3}) in Eq. (\ref{eqd2}) leads to
\begin{equation}
\label{coup2}
\frac{\Sigma_{c}'}{\Sigma_{c}}=\lambda_{c}- 2\epsilon\lambda_{+} \Sigma_{-}\, ,
\end{equation}
and we obtain the differential equation for $\Sigma_{-}$ substituting the above equation together with Eq. (\ref{eqd3}) into Eq.(\ref{eqd1}):
\begin{equation}
\Sigma_{-}'-\epsilon\delta\lambda_{+}(\Sigma_{-})^{2}+\delta\lambda_{c}\Sigma_{-}-\lambda_{-}=0\, .
\label{RiccatiTind}
\end{equation}
This is a first order, quadratic and non-homogeneous ordinary differential equation known as the (time-independent) \textsl{complex Riccati equation}. 
It has unique solution, and can be transformed into an ordinary, homogeneous and second order differential equation with the aid of 
the well-known transformation 
\begin{equation}
\Sigma_{-}=-\frac{1}{\epsilon\delta\lambda_{+}}\frac{u'}{u} \, ,
\label{lambda+final}
\end{equation}
leading to
\begin{equation}
u''+\Gamma u'+\varsigma^{2}u=0\, ,
\label{gener}
\end{equation}
where we defined $\varsigma^2 = \epsilon\delta\lambda_{-}\lambda_{+} \;\;\;\mbox{and}\:\:\:\:\: \Gamma = \delta\lambda_{c} \,$ 
in order to identify it as the classical equation of a damped harmonic oscillator with natural frequency $\varsigma$ and 
damped coefficient $\Gamma$. Its general solution is given by
\begin{equation}
u(\rho)=e^{-\frac{\Gamma}{2} \rho}\left(Ae^{\nu\rho}+Be^{-\nu\rho}\right) \, ,
\label{genersol}
\end{equation}
where $\nu$ is given by Eq. (\ref{eq:nu2}) and constants $A$ and $B$ are  determined from the initial condition $\Sigma_{-}(\rho=0)=0$. 
Using the above results in Eq.(\ref{lambda+final}) we obtain
\begin{equation}
\Sigma_{-}(\rho)=\frac{2\lambda_{-} \sinh(\nu\rho)}{2\nu \cosh(\nu\rho)+\delta\lambda_{c}\sinh(\nu\rho)} \,,\nonumber
\label{lambdamenos}
\end{equation}
which leads to the desired expression written in Eq.(\ref{truesig5}) if we take $\rho=1$. 
Now, using Eq. (\ref{coup2}) and the above result together with the initial condition $\Sigma_{c}(\rho=0)=1$, we can calculate
\begin{equation}
\label{}
\Sigma_{c}=\left(\cosh(\nu\rho)+\frac{\delta\lambda_{c}}{2\nu} \sinh(\nu\rho)\right)^{\frac{2}{\delta}} \, , \nonumber
\label{Lambdac}
\end{equation}
which after taking $\rho = 1$ leads to the desired result of equation in (\ref{truesig4}).
To find $\Sigma_{+}(\rho)$ we replace the above equation in Eq.(\ref{eqd3}) and take into 
account the initial condition $\Sigma_{+}(\rho=0)=0$, obtaining
\begin{equation}
\Sigma_{+}=\frac{2\lambda_{+} \sinh(\nu\rho)}{2\nu \cosh(\nu\rho)+\delta\lambda_{c}\sinh(\nu\rho)} \, ,\nonumber
\label{Lfin}
\end{equation}
which leads to the desired result in Eq. (\ref{truesig5}) with $\rho = 1$.
To finish this section we shall factorize expression (\ref{eq:unfac}) in the usual order:
\begin{equation}
 \hat{G}(\boldsymbol{\Lambda})=e^{\Lambda_{+}\hat{T}_{+}}e^{\ln(\Lambda_{c})\hat{T}_{c}}e^{\Lambda_{-}\hat{T}_{-}} \, .  
\label{eq:factorization2}
\end{equation}
Following a similar process, the correspondent set of coupled differential equations is given by
\begin{eqnarray}
\label{eqd4}
\Lambda_{+}'-\delta\Lambda_{+}\frac{\Lambda_{c}'}{\Lambda_{c}}+\epsilon\delta(\Lambda_{+})^{2}(\Lambda_{c})^{-\delta}\Lambda_{-}'&=&\lambda_{+} \, ,\\
\label{eqd5}
\frac{\Lambda_{c}'}{\Lambda_{c}}-2\epsilon\Lambda_{+}(\Lambda_{c})^{-\delta}\Lambda_{-}'&=&\lambda_{c} \, , \\
\label{eqd6}
(\Lambda_{c})^{-\delta}\Lambda_{-}'&=&\lambda_{-}\, \, .
\end{eqnarray}
Then, solving the above system it is straightforward to show that 
\begin{align}
\label{truej6}
\Lambda_{c}=&\left(\cosh(\nu)-\frac{\delta\lambda_{c}}{2\nu} \sinh(\nu)\right)^{-\frac{2}{\delta}},\\
\label{truej7}
\Lambda_{\pm}=&\frac{2\lambda_{\pm} \sinh(\nu)}{2\nu \cosh(\nu)-\delta\lambda_{c}\sinh(\nu)} \, , 
\end{align}
which are equivalent to Eqs. (\ref{truej4}) and (\ref{truej5}) \cite{DMT-BCH-2020}. It must be noted that our expressions in Eqs. (\ref{truesig4}) and (\ref{truesig5}) differ  by one sign from those found in Ref. \cite{Barnett-1997}  for the special cases of the $\mathfrak{su}(1,1)$ and $\mathfrak{su}(2)$ Lie algebras. However,  we can use Eqs. (\ref{truej6}) and (\ref{truej7}) to check our results as follows. First, notice that the inverse of Eq. (\ref{eq:unfac}) can be 
easily calculated as $\hat{G}^{-1}= \exp{-\lambda_{+}\hat{T}_{+}- \lambda_{c}\hat{T}_{c}-\lambda_{-}\hat{T}_{-}}$, 
\textit{i.e.}, is equivalent to the change $\boldsymbol{\lambda}\rightarrow -\boldsymbol{\lambda}$. 
Using this condition in Eqs. (\ref{truesig4}) and (\ref{truesig5}), it is straightforward to show that 
$\Sigma_{c}\rightarrow (\Lambda_{c})^{-1}$ and $\Sigma_{\pm}\rightarrow -\Lambda_{\pm}$. 
Therefore, from Eq. (\ref{eq:factorization1}) it follows that 
$\hat{G}^{-1}= e^{-\Lambda_{-}\hat{T}_{-}}e^{-\ln(\Lambda_{c})\hat{T}_{c}}e^{-\Lambda_{+}\hat{T}_{+}}$,  
which is consistent with Eq. (\ref{eq:factorization2}), since $\hat{G}\hat{G}^{-1}=\hat{G}^{-1}\hat{G}=1\!\! 1$. 
Notice that the above factorizations do not depend on the unitarity of $\hat{U}_{j}$, but just on the commutation relations 
of the generators of the algebra. 

\end{document}